\documentclass[a4paper,11pt]{article}
\pdfoutput=1 

\usepackage{jcappub} 

\usepackage[T1]{fontenc} 
\usepackage{mathtools}
\usepackage{pgf}
\usepackage[utf8x]{inputenc}
\usepackage{subfig}
\usepackage[toc,page]{appendix}
\usepackage{float}
\usepackage{cleveref}
\usepackage{parskip}

\usepackage[normalem]{ulem}

\newcommand{\diff}{\mathrm{d}}

\title{Preheating in Palatini Higgs inflation on the lattice}

\author[a,1]{F. Dux,\note{Corresponding author.}}
\author[b]{A. Florio,}
\author[c,d]{J. Klari\'c,}
\author[e]{A. Shkerin,}
\author[f,d]{I. Timiryasov}

\affiliation[a]{Laboratory of Astrophysics,\\
  École Polytechnique Fédérale de Lausanne, CH-1015 Lausanne, Switzerland}
\affiliation[b]{Center for Nuclear Theory,\\ Department of Physics and Astronomy,
Stony Brook University, New York 11794-3800, USA}
\affiliation[c]{Centre for Cosmology, Particle Physics and Phenomenology, \\
Université catholique de Louvain, Louvain-la-Neuve B-1348, Belgium}
\affiliation[d]{Laboratory of Particle Physics and Cosmology,\\
École Polytechnique Fédérale de Lausanne, CH-1015 Lausanne, Switzerland}
\affiliation[e]{William I. Fine Theoretical Physics Institute, School of Physics and Astronomy,\\University of Minnesota, Minneapolis, MN 55455, USA}
\affiliation[f]{Niels Bohr Institute, \\
University of Copenhagen, Blegdamsvej 17, DK-2010, Copenhagen, Denmark}

\emailAdd{frederic.dux@epfl.ch}
\emailAdd{adrien.florio@stonybrook.edu}
\emailAdd{juraj.klaric@uclouvain.be}
\emailAdd{ashkerin@umn.edu}
\emailAdd{inar.timiryasov@nbi.ku.dk}

\abstract{
We study preheating following Higgs inflation in the Palatini formulation of gravity. We numerically evolve perturbations of the radial mode of the Higgs field and that of three scalars modeling the gauge bosons. We compare the two non-perturbative mechanisms of growth of excitations---parametric resonance and tachyonic instability---and confirm that the latter plays the dominant role. Our results provide further evidence that preheating in Palatini Higgs inflation happens within a single oscillation of the Higgs field about the bottom of its potential, consistent with the approximation of an instantaneous preheating.
}

\begin{document}
\begin{flushright}
{\small FTPI-MINN-22-08\\ UMN-TH-4117/22 \\
CP3-22-28
}
\end{flushright}
\vspace{-1.cm}
\maketitle
\flushbottom

\section{Introduction}
\label{sec:intro}

Preheating is a period in the evolution of the Universe that connects inflation and the hot Big Bang cosmology. 
During preheating, the energy stored in the inflaton field is transferred to the Standard Model fields and, possibly, dark matter. 
Different perturbative and non-perturbative mechanisms can play roles in this process, see, e.g.,~\cite{Amin:2014eta,Lozanov:2019jxc} for reviews. Preheating parameters, such as its duration and the temperature attained at thermalization, are sensitive to the details of the inflationary scenario and, in particular, to the inflaton's coupling to other species.

One well-motivated candidate for the theory of inflation is  Higgs inflation~\cite{Bezrukov:2007ep}. It stands out among other models in that it does not require new degrees of freedom beyond those of the Standard Model and of General Relativity. 
In this theory, the Higgs field couples to the scalar curvature, and the value of this so-called non-minimal coupling is fixed by observations. 
Since the couplings of the Higgs field to the rest of the Standard Model are known, one can, in principle, compute the preheating parameters and recover the full cosmological history of the Universe. 
The topic of preheating in the model of Higgs inflation has attracted considerable attention, see for instance~\cite{Bezrukov:2008ut,Garcia-Bellido:2008ycs,Enqvist:2013kaa,DeCross:2015uza,Repond:2016sol,Ema:2016dny,DeCross:2016fdz,DeCross:2016cbs,Fu:2017iqg,Sfakianakis:2018lzf,Rubio:2019ypq,He2019,Nguyen:2019kbm,Karam:2020rpa,Hamada:2020kuy,vandeVis:2020qcp,He2021}.

Higgs inflation exists in different incarnations. The original scenario of~\cite{Bezrukov:2007ep} employs the most commonly used metric formulation of General Relativity. In this formulation, all gravitational degrees of freedom are carried by the metric field, and the connection is fixed to be the Levi-Civita one. 
Alternatively, Higgs inflation can be considered in the Palatini formulation of gravity, where the metric and connection are treated as independent variables.\footnote{For other possibilities see, e.g., \cite{Raatikainen:2019qey}, where Higgs inflation was studied in the teleparallel formulation of gravity, and \cite{Azri:2017uor}, where inflation driven by a non-minimally coupled scalar field was studied in the framework of affine gravity. }
Both formulations lead to the usual Einstein field equations of motion in minimally coupled scenarios. However, depending on matter fields and their interaction with gravity, the metric and Palatini formulations can lead to different predictions even when the Lagrangian of the theory is the same.
This is what happens for Higgs inflation and is due to the non-minimal coupling of the Higgs field to gravity. We refer the interested reader to~\cite{Rubio:2018ogq,Tenkanen:2020dge} for reviews of the metric and Palatini Higgs inflation, respectively.
An important prediction of Palatini Higgs inflation is an extremely small tensor-to-scalar ratio~\cite{Rasanen:2017ivk,Enckell:2018kkc}, in contrast to the metric counterpart. Thus, the difference between the two versions of Higgs inflation has observational consequences. 
Another interesting feature of Palatini Higgs inflation is that it has a higher cutoff scale, above which the perturbation theory breaks down, than the metric theory~\cite{Bauer:2010jg}. This allows for a more robust connection between the low-energy observables probed in collider experiments and high-energy inflationary observables~\cite{Shaposhnikov:2020fdv}. 
Next, once fermions are taken into account, the Palatini model naturally generalizes to the Einstein--Cartan framework. The resulting theory shows promising implications, e.g., for inflation and dark matter production~\cite{Langvik:2020nrs,Shaposhnikov:2020gts,Shaposhnikov:2020aen}. All this motivates a closer analysis of preheating in the Palatini Higgs inflation. 

In the metric scenario the main channel of energy transfer from the inflaton field is to the gauge bosons via parametric resonance~\cite{Dolgov:1989us,Traschen:1990sw,Kofman:1994rk,Shtanov:1994ce,Kofman:1997yn,Greene:1997fu}.\footnote{Note that this may lead to the preheating temperature exceeding the cutoff of the theory due to the explosive production of longitudinal component of gauge bosons, see~\cite{DeCross:2016cbs,Ema:2016dny,Gorbunov:2018llf,Bezrukov:2019ylq} and references therein. We comment on this mechanism in section~\ref{ssec:gauge}.}
In contrast, in the Palatini scenario it has been argued in~\cite{Rubio:2019ypq} that the dominant non-perturbative mechanism of preheating is the production of Higgs excitations via a tachyonic instability~\cite{Felder:2000hj,Felder:2001kt}.
The goal of the present paper is to scrutinize this result by performing a fully-fledged numerical simulation that goes beyond the homogeneous approximation of Ref.~\cite{Rubio:2019ypq}.

Specifically, we study the evolution of the Higgs field on an expanding lattice in 3+1 dimensions. For this purpose, we use the recently developed \texttt{$\mathcal{C}osmo\mathcal{L}attice$} package~\cite{Figueroa:2020rrl,Figueroa:2021yhd}. To simplify the problem, we restrict ourselves to the radial degree of freedom of the Higgs field. To model the interaction with the gauge bosons, we introduce three additional scalar degrees of freedom with global couplings to the Higgs field. We also neglect fermions. Under these simplifications, we confirm that the main mechanism driving preheating is the tachyonic instability of the Higgs condensate. The mechanism is very efficient and depletes the condensate within its first period of oscillation. 

In Section~\ref{sec:setup} we introduce the model and discuss the preheating analytically. 
Section~\ref{sec:lattice_study} is dedicated to the lattice studies, and we summarize our conclusions in Section~\ref{sec:conclusion}. 
Finally, several technical details are presented in the appendices.

\section{Setup}\label{sec:setup}

\subsection{The action for inflation}\label{ssec:action}

Let us focus on the part of the theory comprising the Higgs field $H$ and gravity. We adopt the unitary gauge for the Higgs field, $H=(0,h/\sqrt{2})^T$. Then, the action of interest takes the form
\begin{equation}
\label{eq:action1}
        S=\int \diff^{4}x\sqrt{-g}\left[\frac{M_{P}^{2}+\xi h^{2}}{2}g^{\mu\nu}R_{\mu\nu}\left(\Gamma, \partial\,\Gamma\right)-\frac{1}{2}g^{\mu\nu}\partial_{\mu}h\partial_{\nu}h-\frac{\lambda h^{4}}{4}\right]\;.
\end{equation}
Here $M_{P}=2.4\cdot 10^{18}$~GeV is the reduced Planck mass and $\Gamma$ denotes the symmetric connection.
In the Palatini formulation, it is a priori independent of the metric $g_{\mu\nu}$, thus we indicated explicitly that the Ricci tensor is a function of $\Gamma$ and its derivatives. Next, $\xi$ is the non-minimal coupling and $\lambda$ is the Higgs self-coupling. 
We neglect the quadratic term in the Higgs potential in \cref{eq:action1}, since the latter is negligible during inflation and is not relevant for our analysis of preheating.

To analyze the dynamics of the theory (\ref{eq:action1}), it is convenient to get rid of the non-minimal coupling by performing the following Weyl transformation:
\begin{equation}
\label{eq:weyl}
g_{\mu\nu}\mapsto\Omega^{-2}(h)g_{\mu\nu} \;, ~~~ \Omega^2=1+\frac{\xi h^2}{M_P^2} \;.
\end{equation}
At this point one sees the difference between the metric and Palatini formulations: while in the metric version of the theory the rescaling (\ref{eq:weyl}) affects the Ricci tensor, in the Palatini version the latter does not transform since the connection is independent of the metric. The difference manifests itself in the kinetic term of the field $h$ and, ultimately, in the inflationary potential of the canonically normalized scalar field; see, e.g.,~\cite{Rubio:2019ypq,Shaposhnikov:2020fdv} for details. 
The action (\ref{eq:action1}) becomes
\begin{equation}
\label{eq:action2}
S  = \int \diff^{4}x\sqrt{-g}\left[\frac{M_P^2}{2}R-\frac{1}{2}\Omega^{-2}(h)g^{\mu\nu}\partial_{\mu}h\partial_{\nu}h-\Omega^{-4}(h)\frac{\lambda h^{4}}{4}\right] \;.
\end{equation}
This last expression is referred to as the Einstein-frame action, while the original action in \cref{eq:action1} is the Jordan-frame one. 
Since the coupling to gravity is now minimal, the Levi-Civita connection is a solution to the equations of motion.
Therefore, we omitted  $\Gamma$ in the Ricci scalar $R \equiv g^{\mu\nu}R_{\mu\nu}$. 
The next step is to switch to the canonical scalar degree of freedom $\chi$. It is related to $h$ as follows:
\begin{equation}
\label{eq:dchidh}
\frac{\diff\chi}{\diff h}=\frac{1}{\Omega} \;.
\end{equation}
This equation can be solved exactly, yielding
\begin{equation}
\label{eq:h_of_chi}
h(\chi) =\frac{M_{P}}{\sqrt{\xi}}\sinh\left(\frac{\sqrt{\xi}\chi}{M_{P}}\right) \;,
\end{equation}
and we obtain the following action upon substitution
\begin{equation}
\label{eq:action3}
S=\int \diff^{4}x\sqrt{-g}\left[\frac{M_P^2}{2}R-\frac{1}{2}g^{\mu\nu}\partial_{\mu}\chi\partial_{\nu}\chi-V_E(\chi)\right] \;,
\end{equation}
where
\begin{equation}
\label{eq:pot}
V_{E}(\chi)=\frac{\lambda M_{P}^{4}}{4\xi^{2}}\tanh^{4}\left(\frac{\sqrt{\xi}\chi}{M_{P}}\right) \;.
\end{equation}

At large field values, $\chi>M_P/\sqrt{\xi}$, the potential (\ref{eq:pot}) flattens and allows for inflation. The couplings $\xi$ and $\lambda$ at inflationary energy scales are related by the amplitude of primordial spectrum of curvature perturbations~\cite{Planck:2018jri}. We adopt the following values of these parameters.\footnote{Note that this is different from the metric Higgs inflation where typically $\xi\sim 10^3$ \cite{Bezrukov:2007ep}.
The value in \cref{values} lies in the middle of the allowed region $\xi \sim 10^6-10^8$, where the lower limit comes from the possibility of having inflation, and the upper limit comes from the experimental bound on the top Yukawa coupling, see \cite{Shaposhnikov:2020fdv} for details.
}
\begin{equation}\label{values}
\xi=10^7 \;, ~~~ \lambda=10^{-3} \;.
\end{equation}

\subsection{Equation of motion for the inflaton}\label{ssec:eom}
    
In the background specified by the Friedmann–Lemaître–Robertson–Walker (FLRW) metric,
\begin{equation}
\label{ds}
\diff s^2 = - \diff t^2 + a(t)^{2} \diff\mathbf{x}^2 \;,
\end{equation}  
the equation of motion for $\chi$, following from the action (\ref{eq:action3}), reads as follows
\begin{equation}
\label{eq:eom1}
\ddot{\chi}-a^{-2}(t)\nabla^{2}\chi+3\dot{\chi}\mathcal{H}+\frac{\diff V_{E}(\chi)}{\diff\chi}=0 \;.
\end{equation}
Here dot means derivative with respect to $t$ and $\mathcal{H}=\dot{a}/a$ is the Hubble parameter. 
This must be supplemented with the Friedmann equation
\begin{equation}
    \mathcal{H}^2=\frac{\rho}{3M_P^2} \;,
    \label{eq:friedmann1}
\end{equation}
where the energy density $\rho$ associated with the field $\chi$ is
\begin{equation}
    \label{eq:scalardensity}
    \rho = \frac{1}{2}\dot{\chi}^{2}+\frac{1}{2a^{2}}\big(\nabla\chi\big)^{2}+V_E(\chi) \,.
\end{equation}
Consider now a perturbation $\delta\chi(\bold{x},t)$ of the homogeneous background $\chi(t)$, which represents the Higgs field at the onset of preheating. Expanding to the linear order in perturbations, from \cref{eq:eom1} we obtain
\begin{equation}
\label{eq:eom_pert1}
\delta\ddot{\chi}-a(t)^{-2}\nabla^2\delta\chi+3\delta\dot{\chi}\mathcal{H}(t)+\frac{\diff^{2}V_{E}\big(\chi(t)\big)}{\diff\chi(t)^{2}}\delta\chi=0 \;.
\end{equation}
Due to the spatial homogeneity of the background, we can switch to momentum space with
\begin{equation}
\delta\chi(\bold{x},t)=\int\dfrac{\diff^3\bold{k}}{(2\pi)^{3/2}}e^{i\bold{k}\bold{x}}\delta\chi_{\bold{k}}(t) \;.
\end{equation}
The equation of motion for the perturbations then becomes
\begin{equation}
\label{eq:eom_pert2}
\ddot{\delta\chi}_{\bold{k}}
            +3\dot{\delta\chi}_{\bold{k}}\mathcal{H}(t)
            +\Bigg(\frac{\bold{k}^{2}}{a(t)^{2}}
            +\frac{\diff^{2}V_{E}\big(\chi(t)\big)}{\diff\chi(t)^{2}}\Bigg)\delta\chi_{\bold{k}}=0 \;.
\end{equation}
When the term in the parentheses in the above equation is negative for some momentum mode $\bold{k}$, its amplitude
$\delta\chi_\bold{k}$ grows exponentially, signifying tachyonic instability.
The range of momenta for which this can happen is limited; in particular, modes with sufficiently high $\bold{k}$ never suffer from such an instability.

\subsection{Gauge bosons}\label{ssec:gauge}
   
In the metric theory of Higgs inflation, the main mechanism of energy transfer from the Higgs condensate is the non-perturbative production of gauge bosons. The latter, in turn, decay to Standard Model fermions. If the decay rate into fermions is small enough to allow for the gauge bosons to accumulate, the parametric resonance becomes Bose-enhanced and  quickly depletes the energy of the condensate. However, it has been argued in~\cite{Rubio:2019ypq} that in Palatini Higgs inflation, the parametric resonance is subdominant: there, preheating is mainly driven by the tachyonic instability in the Higgs self-interaction, i.e., by the negative coefficient in the $\delta\chi_\bold{k}$-term in eq.~(\ref{eq:eom_pert2}). We are going to test this hypothesis using fully-fledged lattice simulations. 

Including the gauge bosons amounts to promoting the Higgs field derivative in the action (\ref{eq:action1}) to the covariant one.
To simplify the analysis, we will treat the gauge bosons as non-self-interacting scalar degrees of freedom.
More precisely, we introduce the covariant derivative in the Jordan frame, transform it to the Einstein frame, drop all the interaction terms and change the contractions of type $W^\mu W_\mu$ to $W^2$, which is now a scalar degree of freedom.
We are left with a model of 3 massive scalars coupled to the main Higgs field.\footnote{
 This prescription, also utilized in~\cite{Bezrukov:2008ut,Garcia-Bellido:2008ycs,Repond:2016sol}, keeps track of the correct Lorentz/gauge structure of the gauge bosons whose kinetic terms are invariant under scale transformations.
} 
We will refer to these as "scalarized" gauge bosons in the rest of the text.

In summary, the potential associated to the $i^\mathrm{th}$ boson in the Jordan frame is
\begin{equation}
\label{eq:der}
V_J^{i}(h, W_i^\mu) =  \frac{1}{2} m_i^2(h)W^\mu_i W_{i\, \mu}
\end{equation}
where $m_i(h)=g_{(i)} h/2$. 
The couplings $g_{(1,2,3)}$ take the values $g$, $g$ and $g/\cos\theta_W$, correspondingly, with $\theta_W=\tan^{-1}(g^\prime/g)$ the Weinberg angle and $g^\prime$, $g$ the $U(1)_Y$ and $SU(2)$ gauge couplings. 
In this work, we adopt the following values
\begin{equation}\label{values2}
    g^2=0.29 \;, ~~~ \sin^2\theta_W=0.40 \;.
\end{equation}
These are the values at the inflationary energy scale assuming the Standard Model running and consistent with the value of $\lambda$ in \cref{values}.\footnote{More precisely, we take the pole top quark mass $m_t=170.94$ GeV and adopt the central values of the other Standard Model parameters at the weak scale \cite{Zyla:2020zbs}. We then run the RG evolution of $\lambda$, $g$ and $g'$ to the inflationary energy scale $\mu_{\rm inf}$, assuming no new physics. We take $\mu_{\rm inf}=y_t M_P/\sqrt{\xi}=3.3\cdot 10^{14}$ GeV, where $\xi=10^7$ and the top quark Yukawa coupling during inflation is $y_t=0.435$, see \cite{Shaposhnikov:2020fdv} for more details. This way we obtain \cref{values2}. }

Let us comment on the applicability of our approximation of the scalarized gauge bosons. In metric Higgs inflation, it would lead to a significant underestimation of the rate of preheating. Indeed, it has been shown that an explosive production of the longitudinal component of gauge bosons takes place at small field values due to the ``spike'' in the field derivative $\diff\chi/\diff h$~\cite{DeCross:2016cbs,Ema:2016dny}.\footnote{This result should be taken with care since particles produced this way may have energies exceeding the cutoff of the theory~\cite{Bezrukov:2010jz,Gorbunov:2018llf,Bezrukov:2019ylq,Rubio:2019ypq}.} There is not such a spike in the Palatini version of Higgs inflation, and the difference between the transverse and longitudinal components of gauge bosons is irrelevant~\cite{Rubio:2019ypq}. 

In the Einstein frame, the mass term in \cref{eq:der} gets multiplied by the conformal factor $\Omega^{-4}$ from \cref{eq:weyl}. Switching to the canonical field variable according to \cref{eq:dchidh}, we obtain the following potential for the scalarized gauge bosons 
\begin{equation}
\label{eq:gauge_pot}
V_E^i(\chi,W_i)=\frac{g_i^2}{8\xi}M_P^2W_i^2 \tanh^{2}\left(\frac{\sqrt{\xi}\chi}{M_{P}}\right) \;, ~~ i=1,2,3 \;.
\end{equation}
The total Einstein-frame potential for the scalar fields is, therefore, 
\begin{equation}
V_E^{\rm tot}\big(\chi,\big\{ W_i \big\}\big)=V_E(\chi)+\sum_{i=1}^3 V^i_E(\chi,W_i) \;,
\label{eq:totalbosoneinsteinpotential}
\end{equation}
where $V_E(\chi)$ is given in \cref{eq:pot}.
Similarly to \cref{eq:eom1}, we write down the equations of motion of the bosons:
\begin{equation}
    \label{eombosons}
    \ddot{W}_{j}-a(t)^{2}\nabla^{2}W_{j}+3\dot{W}_{j}\mathcal{H}+\frac{\partial V_{E}^{\mathrm{tot}}\big(\chi,\big\{ W_{i}\big\}\big)}{\partial W_{j}}=0\,.
\end{equation}
Finally, the equation of motion for the inflaton~\eqref{eq:eom1} is also modified by replacing $V_E$ with $V_E^{\mathrm{tot}}$.

\section{Lattice study} 
\label{sec:lattice_study}

In this Section, we present results obtained by solving the classical equations of motions using lattice techniques, see~\cite{Figueroa:2020rrl} for an extensive review. For this purpose, we turn to the newly developed open-source \texttt{$\mathcal{C}osmo\mathcal{L}attice$}~\cite{Figueroa:2021yhd}. This package allows us to automatically evolve the canonical equations of motion in the Einstein frame, and is also capable of evolving the full equations of motion of (non)-Abelian gauge degrees of freedom.
While in this work we limit ourselves to scalar degrees of freedom only, the
gauge-invariant discretization handled by \texttt{$\mathcal{C}osmo\mathcal{L}attice$} may greatly simplify the full non-Abelian case which we leave for future work.

\subsection{Lattice units and initial conditions}

For the sake of numerical precision and convenience, we introduce the following dimensionless variables:
\begin{equation}    
\chi= M_{P}\tilde{\chi}\;,\quad\quad W_i = M_{P}\tilde{W}_i\;,\quad\quad t=\frac{2\xi\tilde{t}}{\sqrt{\lambda}M_{P}}\;,\quad\quad \bold{x}=\frac{2\xi\tilde{\bold{x}}}{\sqrt{\lambda}M_{P}} \;.
\label{eq:unitaryglobalrescaling}
\end{equation}
All our simulations are performed in these units. Equation~(\ref{eq:eom1}) becomes
\begin{equation}
    \ddot{\tilde{\chi}}-a^{-2}\tilde{\nabla}^{2}\tilde{\chi}+3\dot{\tilde{\chi}}\tilde{\mathcal{H}}+\frac{\partial\tilde{V}_{E}(\tilde{\chi})}{\partial\tilde{\chi}} =0 \;,
    \label{eq:eomchirescaled}
\end{equation}
where dot now means the derivative with respect to $\tilde{t}$. At this point we recall that if $\tilde{\chi} = \langle \tilde \chi \rangle + \mathcal{O}(\delta \tilde \chi)$ is taken initially to be spatially homogeneous, the term with the Laplacian drops out and the perturbations never grow, yielding an homogeneous version of \cref{eq:eomchirescaled}:
\begin{equation}
    \ddot{\tilde{\chi}}+3\dot{\tilde{\chi}}\tilde{\mathcal{H}}+\frac{\partial\tilde{V}_{E}(\tilde{\chi})}{\partial\tilde{\chi}} =0 \;.
    \label{eq:chiunitaryeomhomogeneousrescaled}
\end{equation}
This equation will be used to specify the initial conditions and for comparison with the lattice simulation.

The total potential in the Einstein frame in terms of the rescaled variables~(\ref{eq:unitaryglobalrescaling}) is 
        \begin{equation}
            \tilde{V}_{E}^{\mathrm{tot}}\big(\tilde{\chi},\{\tilde{W}_{i}\}\big)=\tanh^{4}\left(\sqrt{\xi}\tilde{\chi}\right)+\frac{\xi}{2\lambda}\tanh\left(\sqrt{\xi}\tilde{\chi}\right)^{2}\sum_{i}g_{i}^{2}\tilde{W}_{i}^{2} \; .\label{eq:toteinsteinpotential}
        \end{equation}
For completeness, in appendix \ref{appendix:derivatives} we show the analytical derivatives of this potential with respect to the various fields. 

\begin{figure}[t!]
        \centering
        \input{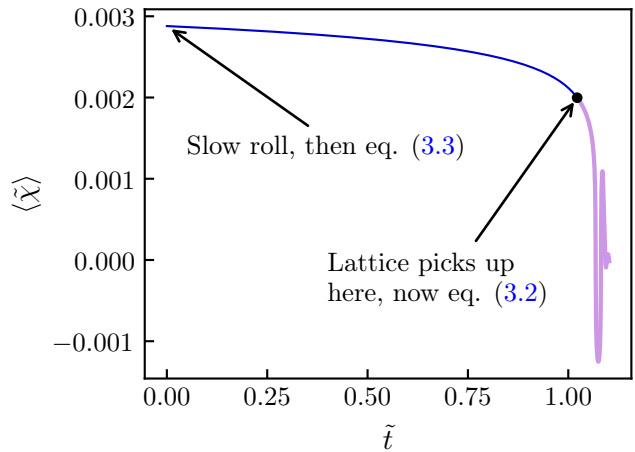}
        \vspace{-0.4cm}\caption{Illustration of how the initial values of $\tilde{\chi}$ and $\dot{\tilde{\chi}}$ are chosen at the beginning of the simulation. First, the homogeneous equation (\ref{eq:chiunitaryeomhomogeneousrescaled}) is integrated from deep inside slow roll. This provides the initial homogeneous components for the evolution of the full partial differential equation (\ref{eq:eomchirescaled}). 
        The y-axis label $\langle \tilde{\chi} \rangle$ denotes the homogeneous field in the first part, and the lattice average calculated by \texttt{$\mathcal{C}osmo\mathcal{L}attice$} in the second part. 
        All units are in terms of (\ref{eq:unitaryglobalrescaling}).
        }
        \label{fig:initialvalues}
\end{figure}

Let us now discuss the initial conditions for the lattice calculation.
To obtain the value of the (homogeneous) inflaton field at the start of the simulation, we evolve it with the ordinary differential equation (\ref{eq:chiunitaryeomhomogeneousrescaled}) from deep within slow-roll. Next, the initial condition for \cref{eq:chiunitaryeomhomogeneousrescaled} itself is dictated by the slow-roll condition:
\begin{equation}
\dot{\tilde{\chi}}_{\rm \,slow \, roll} = - \frac{\tilde{V}'_E(\tilde{\chi})}{3\tilde{\mathcal{H}}} \, .
\end{equation}
At this stage we neglect the gauge bosons since they have vanishing initial abundances.
At a certain moment of time, the values of $\tilde{\chi}$ and $\dot{\tilde{\chi}}$ are passed onto the lattice simulation, where they are used as the initial conditions for the homogeneous component of the field.
The choice of this moment is driven by two considerations.
First, due to the high computational cost of the lattice simulation, it is desirable to evolve the field with \cref{eq:chiunitaryeomhomogeneousrescaled} for as long as possible.
Hence, we do not start the lattice simulation from deep within slow-roll.
As an illustration, figure~\ref{fig:initialvalues} shows the evolution of the (volume averaged) field $\tilde{\chi}$ in the two regimes.
Second, one should be careful not to initialize the lattice with tachyonic modes. From the term in parentheses in \cref{eq:eom_pert2}, such modes are characterised by a limit momentum
\begin{equation}
        \tilde{k}_{\rm tach.} =
        \sqrt{-\frac{\partial^2 \tilde{V}_E}{\partial \tilde{\chi}^2}}
      \approx  \frac{2 \sqrt{2 \xi}}{\cosh{ \sqrt{\xi} \tilde{\chi}}}\;.
\end{equation}
We choose the parameters of the simulation so that modes longer than this threshold are initially absent (see, e.g.,~\cite{Kainulainen:2021eki}). This is achieved by properly setting the initial time (and, hence, the initial value of $\tilde \chi$) and the lattice momentum $\tilde{k}_{IR}$ so that $\tilde{k}_{IR}>\tilde{k}_{\rm tach.}$.%
\footnote{Note that 
there are different ways to handle the initial conditions of the tachyonic modes. For instance, in LATTICEEASY~\cite{FELDER2008929}, all the modes are initialized with the relativistic dispersion relation $\omega_k = |k|$.}

Finally, for the lattice simulation we must take into account the vacuum field fluctuations generated during inflation. Their power spectrum reads as follows
\begin{equation}
        \mathcal{P}_{\delta\tilde{\chi}}(\tilde{k}) = \frac{1}{2a^2 \omega_{\tilde{k},\tilde{\chi}}} \;,\quad \omega_{\tilde{k},\tilde{\chi}} = \sqrt{\tilde{k}^2 + a^2 \frac{\partial^2 \tilde{V}_E}{\partial \tilde{\chi} ^2}} \;.
\end{equation}
We model the said fluctuations by adding a random Gaussian noise with the appropriate variance on top of the homogeneous components of the fields at the onset of the simulation.

\begin{figure}[ht!]
        \centering
        \input{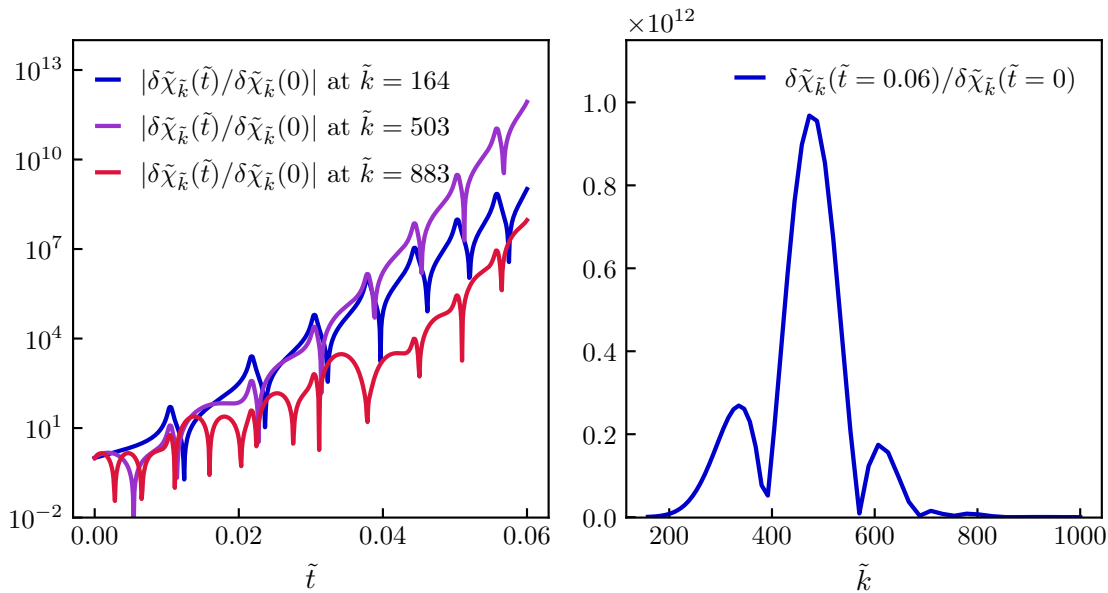}
        \vspace{-0.5cm}
        \caption{ 
        Solution of the linearized equation of motion (\ref{eq:scalarpalatinilinearizedfourier}) during the Palatini-Higgs preheating. 
        (\textit{left}) Evolution of the momentum modes $\delta\tilde{\chi}_{\tilde{k}}(\tilde{t})$
        for several values of $\tilde{k}$. (\textit{right}) Maximum amplitude attained by a mode $\delta\tilde{\chi}_{\tilde{k}}$ after evolving a certain time, for the range of $\tilde{k}$.
        The central peak determines the characteristic momentum scale to keep track of during the simulation which, in turn, determines the appropriate lattice size. The secondary peaks are due to the phase difference between different modes at the time of measurement $\tilde{t}=0.06$. The latter can be chosen arbitrarily, since the growth of perturbations is on average exponential. All units are in terms of (\ref{eq:unitaryglobalrescaling}).
        \label{fig:scalarpalatinimodesthatgrow}}
\end{figure} 

Before moving to the fully-fledged lattice simulation and in order to have an understanding of the relevant energy scales, we start by analyzing the linear evolution of the power spectrum presented in \cref{eq:eom_pert2}. We rewrite it in terms of the units~(\ref{eq:unitaryglobalrescaling}):
\begin{equation}
         \ddot{\delta\tilde{\chi}}_{\tilde{k}}(t)
        +3\dot{\delta\tilde{\chi}}_{\tilde{k}}(t)\mathcal{\tilde{H}}
        +\Bigg(\frac{\tilde{k}^{2}}{a^{2}}
        +\frac{\partial^{2}\tilde{V}_{E}\big(\tilde{\chi}(t)\big)}{\partial\tilde{\chi}(t)^{2}}\Bigg)\delta\tilde{\chi}_{\tilde{k}}(t)=0\label{eq:scalarpalatinilinearizedfourier}
\end{equation}
and solve it numerically. We observe how the amplitude of $\delta\tilde{\chi}_{\tilde{k}}(\tilde{t})$ evolves for a range of momenta $\tilde{k}$. Several examples of such evolutions are presented in the left part of figure~\ref{fig:scalarpalatinimodesthatgrow}. The right part of the same figure shows the spectrum of perturbations resulting from the linear evolution. We see that the dominant modes have momenta concentrated around $\tilde k = 500$: this gives us an idea of the typical momentum that needs to be captured on the lattice. In practice, this is done by setting the physical size $\tilde{L}$ of the simulation box, which is related to the longest mode $\tilde{k}_{\rm IR}$ that the lattice can accommodate:
\begin{equation}
    \label{eq:kir}
    \tilde{k}_{\rm IR} = \frac{2\pi}{\tilde{L}} \, .
\end{equation}
We need values of $\tilde{k}_{\rm IR}$ smaller than 500 so that the relevant modes can be represented on the lattice. At the other end of the spectrum, the shortest possible mode is 
\begin{equation}
    \label{eq:kuv}
    \tilde{k}_{\rm UV} = \frac{\sqrt{3}}{2}N \tilde{k}_{\rm IR} \, ,
\end{equation}
where $N$ is the number of points in one dimension of the lattice. 
For the simulation to be valid, $\tilde{k}_{\rm UV}$ needs to be larger than all the created momenta.
Therefore, we stop the evolution before the modes with $\tilde{k}$ approaching $\tilde{k}_{\rm UV}$ acquire a significant fraction of the total energy.
\begin{figure}[ht!]
        \centering
        \hspace{-0.5cm}
        \input{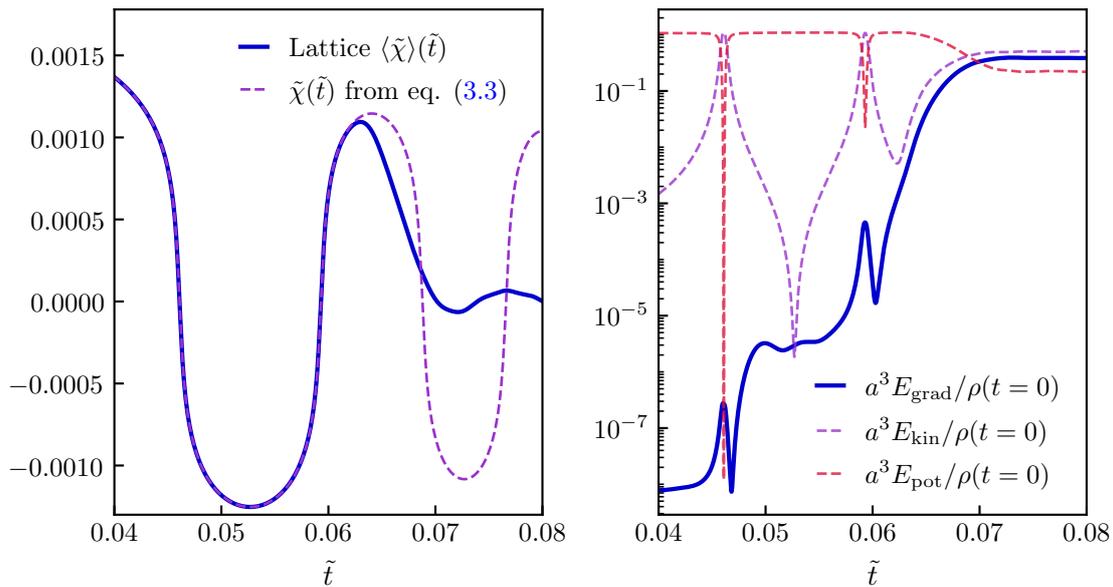}
        \caption{ 
        Early stage of the Palatini--Higgs preheating without gauge bosons.
        (\textit{left}) First oscillations of the (averaged over the lattice volume) inflaton field $\tilde{\chi}$ (solid blue) in comparison with the solution in the absence of perturbations (dashed purple). We see that the back-reaction of the perturbations on the homogeneous mode is almost instantaneous, and the energy is quickly drained off from the inflaton condensate.
        (\textit{right}) Fraction of the total energy of the system
        contained in the gradient, kinetic and potential energies of the Higgs field, as a function of time. All units are in terms
        of (\ref{eq:unitaryglobalrescaling}).}
        \label{fig:scalarpalatinibeginning}
\end{figure}%
\subsection{Lattice results}
\label{subsec:palatinilatticewithoutbosons}
\begin{figure}[ht!]
        \centering
        \input{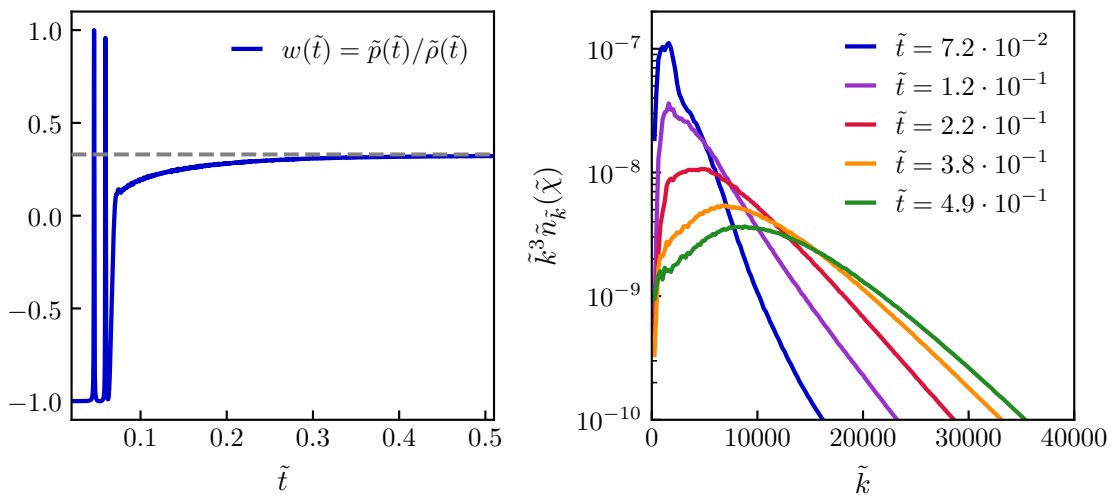}
        \caption{
        Long-term evolution of the Palatini--Higgs preheating without gauge bosons.
        (\textit{left})
          The equation-of-state parameter as a function of time. The plot displays the rate of its convergence to the radiation domination value $1/3$ (the dashed grey line).  
        (\textit{right})
          Energy spectra of excitations of the Higgs field $\tilde{\chi}$ at different moments of time. 
          The blue curve corresponds to the moment at which the spectrum is dominated by the tachyonic excitations created around $\tilde{k}\approx 500$. 
          As time goes by, the peak diffuses towards the UV. We see that during the simulation all modes are well contained within the lattice.
          All units are in terms
        of (\ref{eq:unitaryglobalrescaling}).
        }
        \label{fig:scalarpalatinilongterm}
\end{figure}

\paragraph{Without gauge bosons}
For this simulation, we set $\tilde{k}_{\rm IR}=80$ and $N=512$.
The most striking observation is how quickly the breathing mode of the inflaton is depleted: we see in figure~\ref{fig:scalarpalatinibeginning} that the back-reaction of the perturbations on the homogeneous background is instantaneous, taking less than one oscillation of the inflaton field. 
In the absence of growing perturbations, the field would keep oscillating, following the dashed line in the left part of figure~\ref{fig:scalarpalatinibeginning}. Instead, the second oscillation has an amplitude of just a fraction of the first one.
The right part of the same figure tells us where the energy of the breathing mode went: in the field gradient energy, which violently increased by 8 orders of magnitude in the same time interval.
So at this point, $\tilde{t} \approx 0.065$, the energy density is split roughly evenly between gradient, potential, and kinetic energies.
The Universe is still far from radiation domination: from the left part of figure~\ref{fig:scalarpalatinilongterm} we see that the equation of state parameter $w$ is closer to 0 at $\tilde{t} \approx 0.065$.
This is because only one massive mode is excited.
We see that $w$ approaches $1/3$ not in one period, but in what would be a few dozen periods of the initial oscillations of the inflaton field. There is thus an additional and much longer phase in which other modes become slowly populated. This reflects in the evolution of the energy spectrum, plotted in the right part of figure~\ref{fig:scalarpalatinilongterm}. 
Note that the produced modes fit well within the lattice until the time at which radiation domination is already reached: $\tilde{k}_{\rm UV}$ is large enough such that all the produced modes were well represented by the lattice.
We also check for the consistency of the simulations using the redundancy of the Friedmann equations; see appendix \ref{appendix:energyconservation}.

\paragraph{With gauge bosons}
We now add the gauge bosons to the simulation. A larger lattice size $N$ was required for this purpose, because the gauge bosons tend to migrate further towards the UV than the Higgs excitations. We set $\tilde{k}_{\rm IR}=400$ and $N=648$. 

\begin{figure}[t]
    \centering
    \input{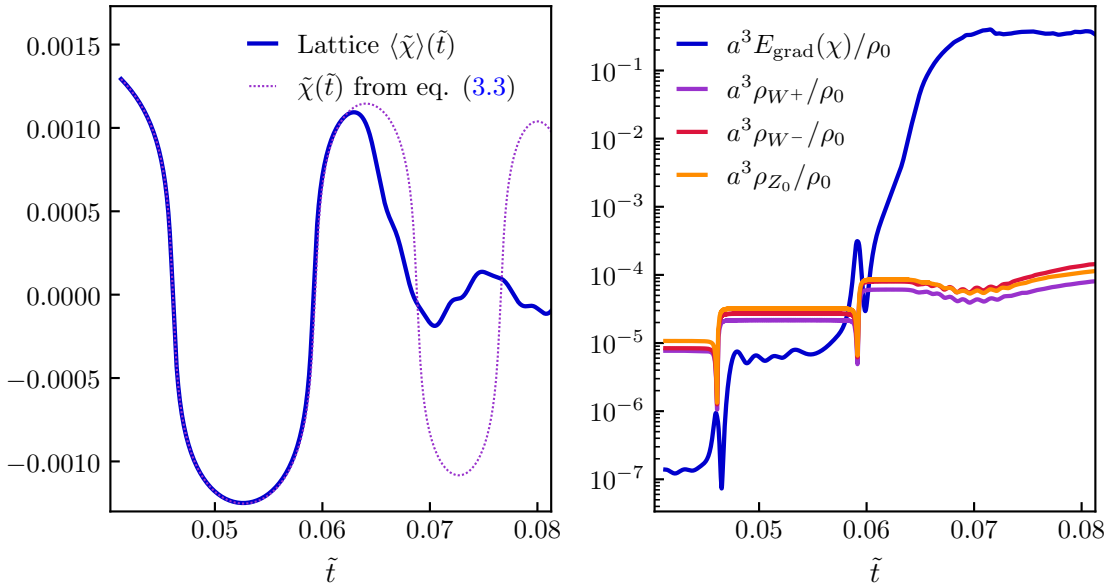}
    \vspace{-0.4cm}
    \caption{Early stage of the Palatini--Higgs preheating with scalarized gauge bosons. 
    (\textit{left})
    First oscillations of the (averaged over the lattice volume) inflaton field $\tilde{\chi}$ (solid blue) in comparison with the solution in the absence of perturbations (dashed purple).
    (\textit{right}) 
    Fraction of the total energy density of the system contained in the gradient of the Higgs field, and the fractions contained in the boson fields. All units are in terms
    of (\ref{eq:unitaryglobalrescaling}). \label{fig:scalarpalatinibosonsfieldandenergies}}
\end{figure}

The behaviour of the Higgs field and of the scale factor is very similar with and without the gauge bosons, because the growth of perturbations in the gauge fields is orders of magnitude below that of the tachyonic excitations. 
This is demonstrated in figure~\ref{fig:scalarpalatinibosonsfieldandenergies}, where on the left we plot the evolution of the inflaton, averaged over the lattice volume, and on the right---the evolution of the relevant components of the energy density.
We furthermore observe the close resemblance of the left parts of figures \ref{fig:scalarpalatinibeginning} and \ref{fig:scalarpalatinibosonsfieldandenergies}. In particular, there is no significant difference in the time at which back-reaction becomes relevant. 

The production of gauge bosons happens through the parametric resonance in the linear regime, more specifically at each zero-crossing of the inflaton field. 
The lattice evolution reproduces this result well, as seen in figure~\ref{fig:scalarpalatinibosonsfieldandenergies}: each zero-crossing of the inflaton field (left panel) causes a brief but rapid growth in the densities of the gauge fields (right panel). Furthermore, the relative increase resulting from each zero-crossing is consistent with that obtained using a Bogoliubov transformation (appendix \ref{appendix:bosonincrease}). 

\begin{figure}[t]
    \centering
    \input{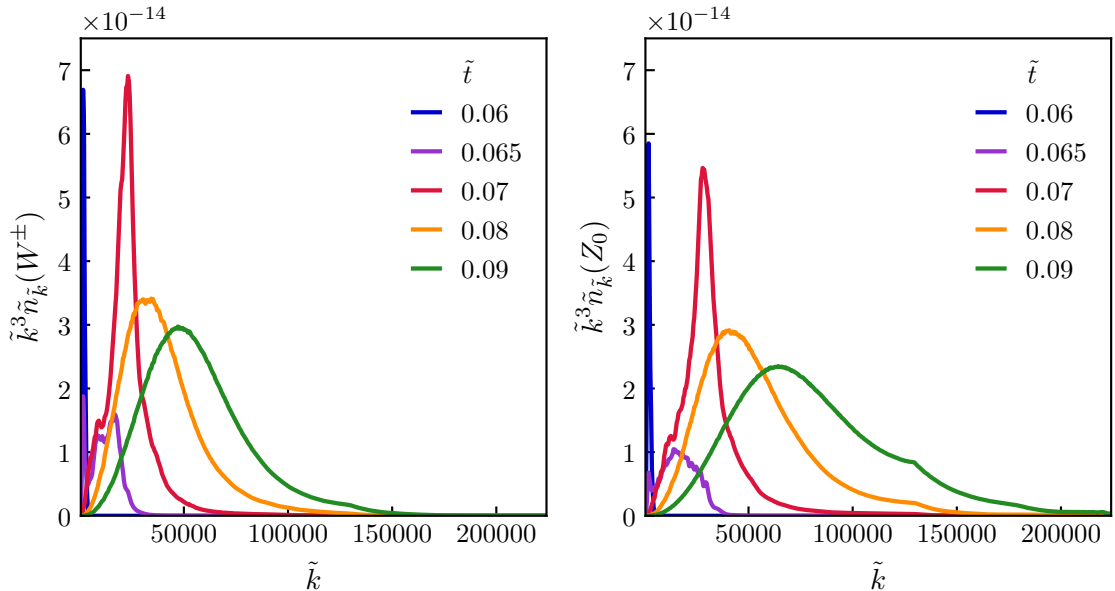}
    \vspace{-0.4cm}
    \caption{
    Energy spectra of excitations in the Palatini--Higgs preheating with scalarized gauge bosons. 
    (\textit{left}) The spectra of $W^+$ and $W^-$ bosons. Here we identify $W_{1,2}$ with $W^\pm$ as it makes no difference in the case of scalarized bosons.
    (\textit{right}) The spectra of the $Z_0$ boson. All quantities are in units of (\ref{eq:unitaryglobalrescaling}).}
    \label{fig:spectrascalarhiggspalatinibosons}
\end{figure}

\paragraph{Energy spectrum of excitations and lattice effects}
We first note that the energy spectra of perturbations of the inflaton field are identical with and without gauge bosons. The spectra of perturbations of the gauge bosons fields are shown in figure~\ref{fig:spectrascalarhiggspalatinibosons}: they are similar to those obtained in Ref.~\cite{Repond:2016sol} in the model of metric Higgs inflation. 

Now the inclusion of gauge bosons required us to use a physically smaller lattice to delay the moment at which the shortest produced mode has a momentum approaching $\tilde{k}_{\rm UV}$. In particular, the gauge coupling of the $Z_0$ boson being the largest one, the diffusion of its perturbations towards the UV is the fastest one and provided the strongest bound on the lattice spacing. 
A direct effect of the physically smaller lattice size is visible when comparing the left panel of figure~\ref{fig:scalarpalatinibosonsfieldandenergies} with that of figure~\ref{fig:scalarpalatinibeginning}: we observe a more pronounced third peak, at $\tilde{t}\sim 0.076$, in the Higgs field average. 
This is an artifact of the smaller physical box size, which prevents the tachyonic production to be as efficient as in the bosonless case.
Let us also mention that the sharp feature in the spectra of gauge bosons, which appears at late times at around $\tilde{k}=\frac{N\cdot \tilde{k}_{IR}}{2}$, is an artifact of radially binning modes distributed on a cube.
The feature is most pronounced for the $Z_0$ boson, as is seen from the right panel of figure~\ref{fig:spectrascalarhiggspalatinibosons}, and indicates the end of validity of our simulation. 
Note, however, that these finite-size effects do not have any measurable consequences for the dynamics of the Universe, since the energy stored in the gauge bosons even at the end of the simulation is $\sim 10^5$ times smaller than the energy of the inflaton excitations.

\subsection{Thermal history and cosmological observables}
\begin{figure}[t!]
       \centering
       \input{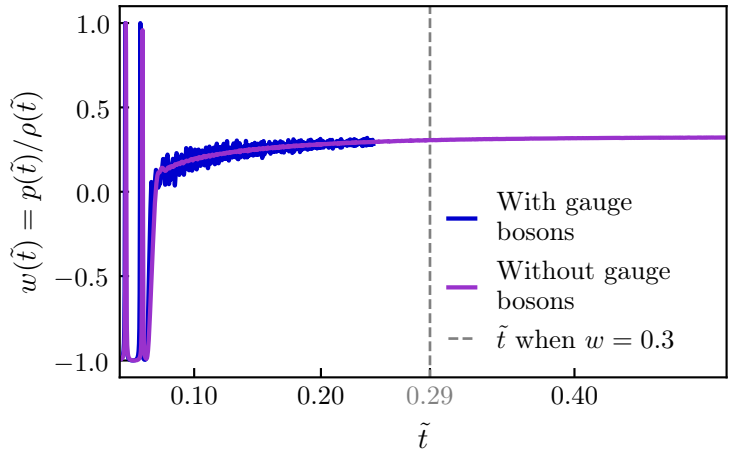} 
        \vspace{-0.4cm}
        \caption{
        Comparison of the evolution of the equation-of-state parameter $w$ in the simulations with (the blue line) and without (the purple line) gauge bosons. 
        The simulation with the bosons terminates earlier
        because of its high computation cost. Clearly, the presence of the bosons does not affect the evolution rate of $w$. The dashed line denotes the time at which $w=0.3$. Time is measured in units of (\ref{eq:unitaryglobalrescaling}). 
        \label{fig:scalarpalatinieqstatewithvswithoutbosons}}
\end{figure}

\begin{figure}[t!]
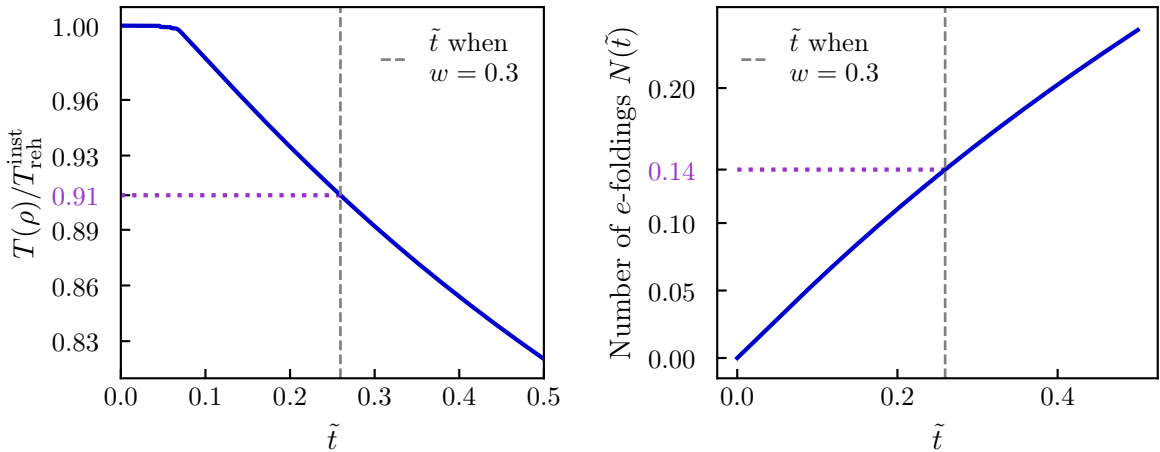

    \hspace{-0.3cm} \input{figs/temperature_progression.pgf}
    \hspace{-0.1cm} \input{figs/scalefactorplot.pgf}
    \vspace{-0.4cm} \caption{
    (\textit{left})
    Cooling down of the Universe as the preheating unrolls. Using the value $w=0.3$ to denote the beginning of the radiation-dominated stage, we obtain the value $0.91\,T_\mathrm{reh}^{\mathrm{inst}}$ of the temperature after preheating, where $T_\mathrm{reh}^{\mathrm{inst}}$ is given in \cref{eq:instantaneous_temp}.
    (\textit{right})
    Logarithm of the scale factor normalized to $1$ at the end of inflation. 
    We see that the entire simulation takes less than one $e$-fold to complete.
    The equation-of-state parameter $w$ reaches $0.3$ by $N_\mathrm{reh} \approx 0.14$.
    We use the data from the bosonless simulation. Time is measured in units of (\ref{eq:unitaryglobalrescaling}).
    }
    \label{fig:temperature_progression}
\end{figure}

Now we discuss the cosmological observables and compare our results to the approximation of an instantaneous preheating.
First we would like to see whether the inclusion of gauge bosons makes the transition to the radiation-dominated Universe faster. Figure~\ref{fig:scalarpalatinieqstatewithvswithoutbosons} shows the evolution of the equation-of-state parameter $w$ with and without the gauge bosons: we observe no qualitative difference between the two cases. Thus, in what follows, we limit our discussion to the bosonless case.
Next, we ask how fast the radiation-dominated stage is reached. Since the exact value $w=1/3$ is only achieved asymptotically, we choose the fairly close value $w=0.3$ as a benchmark point marking the end of preheating. The choice is somewhat arbitrary, but the physical results are not very sensitive to it, as we will see later.       
Using the values from \cref{values,eq:unitaryglobalrescaling} and reading the relevant time off of figure~\ref{fig:scalarpalatinieqstatewithvswithoutbosons}, the duration of preheating is estimated as $t\sim 10^8 M_P^{-1}$. We then conclude that preheating is nearly-instantaneous. 
Finally, we can use the obtained evolution of the energy density between inflation and radiation domination to calculate the cosmological parameters. 
During preheating, the energy content of the Universe is transferred from potential energy of the inflaton field to the $g_*$  effective radiating degrees of freedom.
If all the energy of the Universe is contained in radiation, we can define the ``effective temperature'' as follows:
\begin{equation}
    \label{eq:temp_rho}
    T(\rho)
     =  \left(\frac{30\rho}{\pi^{2}g_{*}}\right)^{1/4} M_P\, .
\end{equation}
Note that in the absence of dissipation, classical simulations do not thermalize to a Boltzmann distribution~\cite[see, e.g.,][]{Micha:2002ey,Berges:2013lsa}.
Nonetheless, the quantity $T(\rho)$ is useful to compare with the results obtained within the approximation of an instantaneous preheating.
If the energy transfer from the inflaton field to the Standard Model degrees of freedom is instantaneous, the potential \eqref{eq:pot} dominates the energy budget, and all of $\rho = \lambda M_P^4 / 4 \xi^2$ is transferred to radiation, giving us the preheating temperature:
\begin{equation}
\label{eq:instantaneous_temp}
    T_\mathrm{reh}^{\mathrm{inst}}
    =
    \left(\frac{30\lambda}{4\pi^{2}\xi^{2}g_{*}}\right)^{1/4}M_{P} \, .
\end{equation}
In the program units this corresponds to $ \tilde{T} = \left(\frac{\xi^2}{\lambda} \frac{120}{\pi^2 g_{*}}\right)^{1/4} \sim 10^4$, which is slightly below the UV cutoff $\tilde{k}_{UV}$ of the lattice simulation.
The simulations give us access to $\rho(t)$, enabling the direct calculation of the effective temperature at any point during preheating. 
This is shown in the left part of figure~\ref{fig:temperature_progression}. The right part of the same figure shows the growth of the scale factor.
So, from the trajectories of the bosonless simulation of figure~\ref{fig:temperature_progression}, we read the following preheating parameters:
\begin{equation}\label{result_without}
    T_{\mathrm{reh}} \approx 0.91 \,T_\mathrm{reh}^{\mathrm{inst}} \;, \quad\quad
    N_{\mathrm{reh}} \approx 0.14\;. 
\end{equation}
Here $N_{\mathrm{reh}}$ is the number of $e$-folds spent during preheating. 
This result agrees with a more detailed linearized analysis of tachyonic preheating where backreaction is taken into account \cite{Tomberg:2021bll}.
This somewhat lower value of $T_{\mathrm{reh}}$ compared to $T_{\mathrm{reh}}^{\mathrm{inst}}$, together with the non-zero $N_{\mathrm{reh}}$,  translate into a slightly lower number of $e$-folds $N_*$ passed between the pivot scale $k_*$ and the end of inflation. 
However, the corresponding correction to the spectral tilt $n_s$ is of order $~10^{-4}$ and is, hence, completely negligible. 
Thereby, we confirm the result of the calculation in Ref.~\cite{Rubio:2019ypq}, made in the linear regime and using an instantaneous preheating.

\section{Conclusions and outlook}
\label{sec:conclusion}

An analytical investigation of preheating in Palatini Higgs inflation showed that it occurs mainly through tachyonic excitations of the Higgs field~\cite{Rubio:2019ypq}. 
In the present paper, we confirmed this result by performing a fully-fledged lattice calculation of the nonlinear evolution of the field perturbations. 
We did this in a model comprising the radial degree of freedom of the Higgs field and three scalarized, non-interacting gauge bosons. 
We found that the inclusion of such bosons does not significantly affect the parameters of preheating, such as its duration and temperature.
Specifically, we found that the pressure-to-density ratio approaches $0.3$ within $\sim 0.1$ $e$-fold, and that the equivalent temperature is roughly 90\% of that of an instantaneous preheating. 
The deviation from the instantaneous limit results in the shift of the spectral index by the negligible amount $\sim10^{-4}$.

It would be interesting to study the impact of other Standard Model degrees of freedom on preheating. 
For example, in the metric version of Higgs inflation, the mechanisms other than the tachyonic instability can lead to the violent preheating: the issue of the spike in the mass of the longitudinal gauge boson is one, but also, for instance, the violent excitation of the phase of an Abelian Higgs model~\cite{Ema:2016dny}. 
Simulations incorporating the full $SU(2)$ structure would shed light on the relevance of such phenomena for the preheating in Palatini Higgs inflation.

Another direction for future research is the inclusion of the coupling to metric perturbations. 
This would allow the study of the production of gravitational waves during preheating, which could be interesting for phenomenology~\cite[see, e.g.,][]{Koivunen:2022mem}.

\section*{Acknowledgements}
 We thank Eduardo Grossi, Javier Rubio, Mikhail Shaposhnikov and Anna Tokareva for the helpful discussions and comments.
 We also thank the anonymous referee for their helpful
comments and suggestions.
 The computations were performed at University of Geneva on the "Yggdrasil" and "LESTA" HPC clusters. 
 The work of F.D. was supported by his thesis supervisor, Frederic Courbin. 
 The work of A.S. was partially supported by the Department of Energy Grant DE-SC0011842.
 The work of I.T. was partially supported by ERC-AdG-2015 grant 694896, by the Swiss National Science Foundation Excellence grant 200020B\underline{ }182864,  by the Carlsberg foundation, and by
 the European Union’s Horizon 2020 research and innovation program under the Marie Sklodowska-Curie grant agreement No. 847523 `INTERACTIONS'.
 J.K. acknowledges the support of the Fonds de la Recherche Scientifique - FNRS under Grant No. 4.4512.10.
 A.F. is supported by the U.S. Department of Energy, Office of Science, Office of Nuclear Physics, grants Nos. DE-FG-02-08ER41450.

\appendix

\section{Derivatives of the Einstein frame potential}
\label{appendix:derivatives}

Here we write the derivatives of potential (\ref{eq:toteinsteinpotential}), in the rescaled units (\ref{eq:chiunitaryeomhomogeneousrescaled}). The derivatives with respect to the inflaton field read
\begin{equation}
    \frac{\partial \tilde{V}_{E}^{\mathrm{tot}}}{ \partial \tilde{\chi}}=\frac{\text{\ensuremath{\sqrt{\xi}}}}{\lambda}\frac{\tanh\left(\sqrt{\xi}\tilde{\chi}\right)}{\cosh^{2}\left(\sqrt{\xi}\tilde{\chi}\right)}\Bigg(\xi\sum_{i}g_{i}^{2}\tilde{W}_{i}^{2}+4\lambda\tanh^{2}\left(\sqrt{\xi}\tilde{\chi}\right)\Bigg) \;,
\end{equation}
\begin{multline}
\frac{\partial^{2}\tilde{V}_{E}^{\mathrm{tot}}}{\partial \tilde{\chi}^{2}}=\frac{1}{\lambda}\frac{\text{\ensuremath{\xi}}}{\cosh^{4}\left(\sqrt{\xi}\tilde{\chi}\right)}\Bigg[\xi\Big(2-\cosh\big(2\sqrt{\xi}\tilde{\chi}\big)\Big)\sum_{i}g_{i}^{2}\tilde{W}_{i}^{2}\\+4\lambda\Big(4-\cosh\big(2\sqrt{\xi}\tilde{\chi}\big)\Big)\tanh^{2}\left(\sqrt{\xi}\tilde{\chi}\right)\Bigg] \;.\label{eq:appendixpalatinid2vdchi2}
\end{multline}
The derivatives with respect to the gauge fields read
\begin{equation}
\frac{\partial \tilde{V}_{E}^{\mathrm{tot}}}{\partial \tilde{W_{j}}}=\frac{\xi}{\lambda}\tanh^{2}\left(\sqrt{\xi}\tilde{\chi}\right)g_{j}^{2}\tilde{W}_{j} \; ,\label{eq:appendixpalatinidvdw}
\end{equation}
\begin{equation}
\frac{\partial^{2}\tilde{V}_{E}^{\mathrm{tot}}}{\partial\tilde{W}_{j}^{2}}=\frac{\xi}{\lambda}\tanh^{2}\left(\sqrt{\xi}\tilde{\chi}\right) g_{j}^{2} \; .\label{eq:appendixpalatinid2vdw2}
\end{equation}
The files implementing the derivatives can be found in a GitHub repository~\cite{githubrepo}.

\section{Conservation of energy in the simulations}
\label{appendix:energyconservation}

\begin{figure}
    \centering
    \input{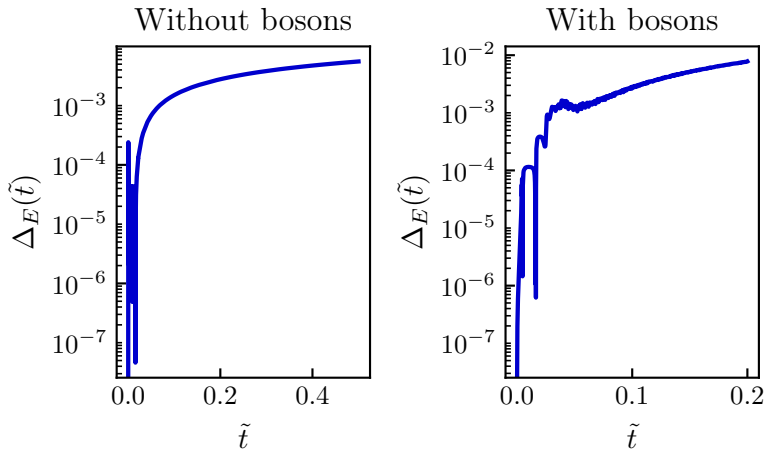}
    \vspace{-0.4cm}\caption{Relative energy conservation of both simulations, with and without gauge scalars. All units are in terms of (\ref{eq:unitaryglobalrescaling}).
    \label{fig:energyconservation}}
\end{figure}

We use the first Friedmann equation (\ref{eq:friedmann1}) to evolve the scale factor in the simulations. The second Friedmann equation,
\begin{equation}
    \label{eq:friedmann2}
    \frac{\ddot{a}}{a}=-\frac{1}{6M_{P}^{2}}\bigg(\rho+3p\bigg) \;,
\end{equation}
can be used to check the consistency of the simulation at any point in time. In the FRLW background~(\ref{ds}), the energy density $\rho$ is given in \cref{eq:scalardensity} and the pressure is given by
\begin{equation}
    p = \frac{1}{2}\dot{\chi}^{2}-\frac{1}{6a^{2}}(\nabla\chi)^{2}-V_E(\chi) \,.
\end{equation}
During the simulation, we compute the left hand side (LHS) and the right hand side (RHS) of \cref{eq:friedmann2} and use them to estimate the relative error
\begin{equation}
    \label{eq:relenergyconservation}
    \Delta_{E}(\tilde{t})=\bigg\vert\frac{{\rm LHS}-{\rm RHS}}{{\rm LHS}}\bigg\vert \;.
\end{equation}
The result is shown in figure~\ref{fig:energyconservation}, both in the absence and presence of the gauge bosons.

Overall the energy is conserved to less than 1\%, similar to what was attained in~\cite{Repond:2016sol}.

\section{Increase of the energy density contained in the bosons at each zero-crossing}
\label{appendix:bosonincrease}
We start with the phase averaged increase in the number density
of the gauge bosons after the $j^{\rm th}$ zero-crossing \cite{Kofman:1997yn, Rubio:2019ypq}:
\begin{equation}
\bigg(\frac{1}{2}+n_{\tilde{k}}^{j+1}\bigg)=\bigg(1+2C(t_{j})\bigg)\bigg(\frac{1}{2}+n_{\tilde{k}}^{j}\bigg)\;.
\end{equation}
 For convenience, we unpack the $C(t_j)$ term:
\begin{equation}
n_{k}^{j+1}=n_{k}^{j}+\big(1+2n_{k}^{j}\big)\exp\bigg(-\sqrt{\frac{2}{3}}\pi\frac{k^{2}}{g\, a^{2}(t_{j})\mathcal{H}(t_{j})M_{p}}\bigg)\,\,.
\end{equation}
 The total relative increase $I(j)$ in the energy density after the
$j^{{\rm th}}$ zero-crossing is thereby the ratio
\begin{equation}
I(j)=\frac{\int_{0}^{\infty} dk\,\,k^{3}\, n_{k}^{j+1}}{\int_{0}^{\infty} dk\,\,k^{3}\, n_{k}^{j}}\,\,.\label{eq:ratio_of_gauge_boson_number_density}
\end{equation}
We introduce two factors $B$ and $B'$ converting $n_{k}$ to our
units 
\begin{align}
n_{k} & =B\tilde{n}_{\tilde{k}},\\
dk & =B'd\tilde{k},
\end{align}
these are readily eliminated from the ratio:
\begin{equation}
I(j)=\frac{\int_{0}^{\infty}d\tilde{k}\,\,\tilde{k}^{3}\tilde{n}_{\tilde{k}}^{j+1}}{\int_{0}^{\infty}d\tilde{k}\,\,\tilde{k}^{3}\tilde{n}_{\tilde{k}}^{j}}\,\,.
\end{equation}
Next, we take the initial distribution of $\tilde{n}_{\tilde{k}}$
on our lattice, that is a step function 
\begin{equation}
\tilde{n}_{\tilde{k}}(0)=A\bigg(\Theta(\tilde{k})-\Theta(\tilde{k}-\tilde{k}_{{\rm cutoff}})\bigg)
\end{equation}
 with $A\sim10^{16}$ and $\tilde{k}_{{\rm cutoff}}=1500$. Upon numerical integration of (\ref{eq:ratio_of_gauge_boson_number_density}),
for the $W^\pm$ bosons, we obtain the following ratio:
\begin{equation}
I(1)\sim I(2)=2.895(5)
\end{equation}
which can be compared to the ratio obtained on the lattice:
\begin{equation}
I_{\rm lattice}(1)\sim I_{\rm lattice}(2)=2.9\pm0.1
\end{equation}
Where the uncertainty is stemming from the somewhat arbitrary times ("just before" and "just after" a zero-crossing) at which the densities are read from the lattice trajectories. 

\addtocontents{toc}

\bibliographystyle{JHEP}
\bibliography{refs}

\end{document}